\documentclass[usenatbib]{mn2e}
\usepackage{graphicx}
\graphicspath{{./images/}}
\usepackage{tabularx}
\usepackage{amsmath}
\newcommand{\beq}{\begin{equation}}
\newcommand{\eeq}{\end{equation}}

\title[Using Kinematic Properties of Pre-Planetary Nebulae to Constrain  Engine  Paradigms]{Using Kinematic Properties of Pre-Planetary Nebulae to Constrain  Engine  Paradigms}

\author[ Blackman, Lucchini]{Eric G. Blackman$^{1}$, Scott Lucchini$^{1}$\\$^{1}$Department of Physics and Astronomy, University of Rochester, Rochester, NY 14627, USA}
\begin{document}

\date{}

\pagerange{\pageref{firstpage}--\pageref{lastpage}} \pubyear{2011}

\maketitle

\label{firstpage}

\begin{abstract}
Some combination of binary interactions and accretion plausibly conspire to produce the ubiquitous  collimated outflows from planetary nebulae (PN) and their presumed pre-planetary nebulae (PPN) precursors. But which  accretion engines are viable?  The difficulty in observationally  resolving the engines warrants  the pursuit of indirect constraints. We show  how   kinematic outflow data for 19  PPN can be used to determine the minimum required  accretion rates.  We consider main sequence (MS) and white dwarf (WD) accretors and five example accretion rates   inferred from published models to compare with the  minima derived from outflow momentum conservation. While our primary goal is to show the  method in anticipation of more data and better  theoretical  constraints, taking the present results at face value already rule out  modes of accretion:  Bondi-Hoyle Lyttleton (BHL) wind accretion and wind Roche lobe overflow (M-WRLOF, based on Mira parameters)    are too feeble for all 19/19 objects  for  a MS accretor.    For a WD accretor, BHL is ruled out for 18/19 objects and M-WRLOF for 15/19 objects.  Roche lobe overflow (RLOF) from the primary at the Red Rectangle level
 can  accommodate 7/19 objects, though RLOF modes  with higher accretion rates are not yet ruled out.
 Accretion modes operating from within common envelope evolution can  accommodate all 19 objects,  if jet collimation can be maintained. Overall,  sub-Eddington rates for a MS accretor are acceptable but 8/19 would  require super-Eddington rates for a WD.

\end{abstract}


\section{Introduction}
\label{sec:intro}

 . 

 Binary paradigms that involve accretion   \citep{soker94,soker96,soker97,soker98,reyes99,blackman01,2006MNRAS.370.2004N} are plausibly fundamental   to producing many of the asymmetric outflows  observed in planetary nebulae (PN) and pre-planetary nebulae (PPN) \citep{balick02}.  Their ubiquity is statistically consistent with the frequency of binaries  \citep{demarco11,demarco13}. The influence of the latter also need not even imply their present presence if, for example, tidal shredding is involved \citep{2006MNRAS.370.2004N,nordhaus10}.

Binaries  provide a source of angular momentum and  free energy to form accretion disks and  such disks can in turn amplify magnetic fields  that transport  angular momentum  locally and on large scales.  The  latter may manifest as bipolar  jets and/or winds,  If  jetted  PN and PPN involve such disks, then both the birth and death of  stars represent simllar highly aspherical states that sandwich the more  spherical
life of  stars on the main sequence (MS).

PPN are  distinguished from  PN in that the former are reflection nebulae and the latter are ionization nebulae , and it is
likely that PPN  transition to the latter  when the central star of the primary becomes sufficiently exposed to ionize the ambient nebular gas  \citep{2000oepn.book.....K} . Observations of  high momentum outflows from PPN  \citep{ppne,sahai08} are thus  important for understanding the engines of both PPN and PN.    
 Even with a relatively conservative allotment for uncertainties,  \cite{ppne} concluded that   ~80\% of 28 sources detected with seemingly bipolar CO molecular outflows  had  scalar momenta  in excess of that which could be  be supplied
by radiation.  The kinematic requirements of PPN are  more demanding than those of PN, but  if PPN evolve to PN, then  constraints on engine paradigms of PPN also constrain PN engines.

If binaries and accretion are important to  PPN and  PN,  a next question is   which binary accretion scenarios are viable? Candidate scenarios  include various modes of  accretion onto either MS or white dwarf (WD) companions, or accretion from a  shredded secondary onto a primary core. But the observational difficulties of detecting  binaries  and accretion disks warrant  indirect constraints.    Outflow observations can provide constraints on the needed momentum,  energy, and accretion rate.   \cite{huggins} estimated the kinetic energy  content of  jets and tori from some PPN and found that their sum was greater  than the binding energy of the envelope but less than the available energy if the primary envelope accreted onto a main sequence companion. However (despite its paper title)  \citet{huggins}  did not  constrain the {\it power} or {accretion rates}. Determining the minimum required accretion rates is fundamental for assessing the viability of an engine scenario.  Here we   give a method  for doing this and to apply it to known PPN where kinematic data are
available.  In the long run, the method awaits  more data and theory but we  find that  some modes of accretion can be already  be eliminated for the published data used.

In Sec. 2 we discuss how to use  observed outflow momenta and mass loss rates to constrain  minimum accretion rates at the engine.
In Sec. 3 we discuss 5 modes of accretion for which theoretical calculations or simulations  have provided an accretion rate. 
In Sec. 4 we discuss the published data   and
combine this data with the results of sections 2 and 3 to produce constraint plots. We discuss the implications of this plots with respect to  ruling out
some modes of accretion.
 We conclude in section 5.




\section{Minimum required accretion rates}
\label{sec:eqs}

Regardless of the particular accretion mechanism,
all jets formed from an accretion disk but have  a mechanical luminosity  less than that of the rate of energy supply to the engine.
The latter is given by \citep{2002apa..book.....F}
$\frac{1}{2}\frac{G M_a\dot{M_a}}{R_a}$, where $M_a$ is the mass of the central accretor, $R_a$ is the inner radius of the disk, $G$ is the gravitational constant, and $\dot{M_a}$ is the accretion rate,
The  jet mechanical luminosity then satisfies 
\begin{equation}
L_{mec}= {1\over 2} {\dot M}_{\textrm{j}} v_{j,N}^2 \le \frac{1}{2}\frac{G M_a\dot{M_a}}{R_a}= {1\over 2} {\dot M}_{\textrm{a}} v_k^2(R_{\textrm{a}}),
\label{1}
\end{equation}
where  ${\dot M}_{\textrm{j}}$ is the jet mass loss rate, $v_k(R_{\textrm{a}})=\left({GM_a\over R_a}\right)^{1/2}$ is the Keplerian speed at the inner disk radius and the "naked"
 jet speed  
\beq
v_{j,N}\equiv  Q v_k(R_{\textrm{a}})  =520 {\rm km/s} \left({Q\over{\sqrt 2} }\right)\left({M_a\over M_\odot }\right)\left({R_a\over R_\odot }\right)^{-1}
\label{0.1}\eeq
 is the maximum speed the jet would have  before
slowing down from swept up material.  Here $Q$ is a numerical factor typically satisfying $1< Q <5$ in  jet models \citep{blandford82,pelletier92,lb03,ferrari09}, but it can depend on the environment into which the flow propagates, the location of the Alfv\'en surface and whether the model describes a steady magneto-centrifugal launch or a magnetic tower.  In the PPN/PN context, the jet  propagates into an envelope of material,
and the scales at which the jet velocities are measured for PPN are large
compared to those of engine launch.  

Eq. (\ref{1}) and Eq. (\ref{0.1})  imply
\beq
\dot{M_a}\ge Q^2\dot{M_j}.
\label{0.3}
\eeq

To use Eq. (\ref{0.3}) to constrain engine  models for a given choice of $Q$,
 we need an expression for $\dot{M_j}$ in terms of 
observables. 
We first write 
 \beq
  {{\dot M}_j\simeq M_{j,N} / t_{\textrm{acc}}},
 \label{0.5}
 \eeq 
where $M_{j,N}$ is the mass ejected the naked jet without swept up material,  
$ t_{\textrm{acc}}$ is the acceleration time scale of the jet for which an upper limit is observable (discussed later).
As the observed outflow is contaminated by  swept up  material, 
  inferences about  ${\dot M}_j$ must account for this using momentum conservation
  \beq
 p_j=M_{\textrm{j,ob}} v_{\textrm{j,ob}}=M_{j,N} v_{j,N},
 \label{0.4}
\eeq
where $ M_{\textrm{j,ob}}$ is the mass observed in the outflow, 
 $v_{\textrm{j,ob}}$ is the observed jet velocity. 
 Plugging Eq.  (\ref{0.1}) into  Eq. (\ref{0.4}) and the result into (\ref{0.5}), Eq. (\ref{0.3}) then  gives
 \beq
 \begin{array}{r}
{ \dot M}_a\ge  10^{-4}\left({Q \over 2}\right)\left({{M}_a\over {M}_\odot }\right)^{-{1\over 2}}\left({R_a\over {R}_\odot }\right)^{1\over 2}\\
\times\left({M_{\textrm{j,ob}}\over 0.1 {\dot M}_\odot }\right)\left({{v}_{\textrm{j,ob}}\over 100{\rm {km/ s}}}\right)\left({{t}_{acc}\over 500{\rm yr}}\right)^{-1}.
 \label{0.7}
 \end{array}
 \eeq

We will use Eq. (\ref{0.7}) for both MS and WD stars. Assuming that the inner disk radius equals that of the stellar photosphere (ignoring the effect from a magnetosphere). for low mass MS stars of radius $R_*$, the mass radius relation  is approximately \citep{1991Ap&SS.181..313D}
$
R_a \sim 0.99 R_*\simeq R_\odot \left({M_a\over M_\odot}\right)^{0.89},  
$ for zero age MS and $
R_a \sim 2 R_*\simeq R_\odot \left({M_a\over M_\odot}\right)^{0.75},  
$ for terminal age MS. 
Thus the right side of  Eq. (\ref{0.7}) will 
  only weakly decrease with increasing mass.

 Eq. (\ref{0.7}) gives lower limits on ${\dot M}_{a}$ that are a factor ${v_{j,N}\over v_{j,ob}}>1$, larger than that 
 which would arise if we had followed the same procedure using energy conservation instead of Eq. (\ref{0.4}). Thus using momentum conservation  is essential to obtain the more stringent limit. 
In addition,  bulk flow energy can be lost  via radiation in the outflow or conversion of bulk to thermal energy.   Nevertheless, the accretion rates  are still minima since (i) they presume all of the accreted 
 power goes into the outflow; (ii) the observed values used for   $t_{\textrm{acc}}$ are generally upper limits; and  (iii) assumptions  and uncertainties in the interpretation of the CO lines  scalar momenta are underestimates \citep{ppne}.



\section{Theoertical Accretion Rates}

\label{sec:summary}

\subsection{Bondi-Hoyle-Littleton (BHL)} 
\label{sec:bhl}
For a primary AGB wind emitter of order $1M_\odot$ the  radius above which the flow is radiatively accelerated (via dust) to its 
steady wind speed   is $\sim 10$AU  \citep{2008MNRAS.385..215S}. 
For a binary  with such a  primary interacting with an accreting secondary located outside
this radiative acceleration radius $r_{dust}$, the  BHL
 model \citep{2004NewAR..48..843E,xraybook} provides a good approximation to the accretion rate, namely
\begin{equation}
\dot{M}_{\textrm{BH}}=10^{-9}{\rm M_\odot/{\rm yr}}\left({\dot{M}_{w}\over10^{-5}{\rm M_\odot/{\rm yr}}}\right)
\left({M_{S}\over M_{P}}\right)^2 \frac{\left(\frac{v_{\textrm{or}}/v_{w}}{0.1}\right)^4}{\left[1+\left({v_{\textrm{or}}\over v_{w}}\right)^2\right]^{\frac{3}{2}}},
\label{BH}
\end{equation}
where  $\dot{M}_{w}$ is the mass loss rate of the primary star,  
$M_{S}$ is the mass of the secondary, $M_{P}$ is the mass of the primary, $v_{w}$ is the velocity of the primary's wind \citep{xraybook},
$v_{\textrm{or}}=\left[{M_P/M_S\over 1+ M_P/ M_S}{G M_P\over a_{\textrm{or}}}\right]^{1/2}$ is the (circular) orbit speed of the secondary for an orbital separation $a_{\textrm{or}}$ about the center of mass. 
For $M_P=1.5M_\odot$, $v_w=10$km/s, $M_S=1M_\odot$, and $a_{\textrm{or}}=10$AU,  
${\dot M}_{\textrm{BH}}=1.1\times 10^{-6}M_\odot$/yr, if ${\dot M}_w=1.3 \times 10^{-4}M_\odot$/yr.
Numerical simulations of \cite{martin}, generally confirm the
utility of Eq. (\ref{BH}).

\subsection{Wind Roche-Lobe Overflow  (WRLOF)}
\label{sec:wrlof}
When $a_{\textrm{or}}$ is  large enough such that the Roche lobe of the primary is outside of the primary's radius but less than $r_{dust}$,   the wind can fill the primary's Roche lobe  and  overflow onto the secondary.
This was first discussed and simulated  by \citet{mohamed} for  Mira ( hearafter M-WRLOF). 
For   ${\dot M}_w=2 \times 10^{-5} M_\odot$/yr,  $M_P= 1 M_\odot$,  $M_S =0.6 M_\odot$,  $a_{\textrm{or}}=20$AU, $r_{dust} = 6R_P = 10$AU, and the Roche radius of the primary $R_{L,P}=8.5$AU, 
 they found a M-WRLOF accretion rate of  ${\dot M}_{\textrm{WR}}\sim 0.5{\dot M}_w\simeq 5\times 10^{-7}\;\mathrm{M_{\sun}\: yr^{-1}}$. 
    This is $\sim20$ times the BHL rate for this set of parameters using Eq. (\ref{BH}).
   

\subsection{Common Envelope}
\label{sec:ce}
 \citet{rt} simulated a  common envelope (CE) binary evolution
  using  $M_P=1.05M_\odot$  (with core of mass $0.36M_\odot$) and  $M_S= 0.6M_\odot$. 
  Initially,  $a_{\textrm{or}}=4.3\times 10^{12}$cm  and an initial {\it red giant} primary radius of $2.3 \times 10^{12}$cm.   The evolution was followed for 56.7 days of simulation time ($\sim 5$ orbits) by which time $a_{\textrm{or}}$ shrunk by a factor of $7$.   The  average accretion rate onto the secondary over the duration of the simulation   and onto the primary core were
   ${\dot M}_{\textrm{CE}}
   \sim 10^{-2}M_\odot$/yr and ${\dot M}_{\textrm{CE,P}}\sim 6\times 10^{-2}M_\odot$ respectively.
  %


These rates are significantly greater than the Eddington accretion rates for a WD or MS star. The Edddington accretion rate is that 
which produces the Eddington luminosity. The latter is given by 
\begin{equation*}
L_{\textrm{ed}}=\frac{1}{2}\frac{G M_{\textrm{a}} \dot{M}_{\textrm{ed}}}{R_{\textrm{a}}}=1.23\times 10^{38}\: (M_{\textrm{a}}/M_{\sun})\; \mathrm{erg/s}
\end{equation*}
so that the Eddington accretion rate is 
\begin{equation}
\label{eq:eddington} 
\dot{M}_{\textrm{ed}}=\frac{2 L_{\textrm{ed}} R_{\textrm{a}}}{G M_{\textrm{a}}}=2.9\times 10^{-5}\left(R_{\textrm{a}}\over 10^9{\rm cm}\right)\; \mathrm{M_{\sun}\: yr^{-1}},
\end{equation}
where  $L_{\textrm{ed}}=1.23\times 10^{38}\: (M_{\textrm{a}}/M_{\sun})\; \mathrm{erg/s}$, 
where we have scaled to the radius of a WD.
This Eddington rate for a WD is included as a horizontal gridline in Figure \ref{fig:mdotaccmomentum}a and 
for a MS star of radius $R_{\textrm{a}} =7\times 10^{10}$cm in Figure \ref{fig:mdotaccmomentum}b.

Note  that accretion onto the {\it primary} via tidal shredding of the secondary \citep{2006MNRAS.370.2004N,Nordhaus11} of a low mass star or
large planet could be super-Eddington. Similar to the case of black holes  \citep{abramowicz1980},   it  may here too proceed with an optically
and  geometrically thick disk in which the radiative diffusion time is  slow compared to the accretion time, and bipolar outflows of super-Eddington mechanical luminosity. More work for this mode of accretion around white dwarfs is thus desirable.

\subsection{Red-Rectangle Roche lobe overflow (RR-RLOF)}
\citet{wittrr} report on observations and  detailed modeling of the Red Rectangle (RR)  PPN. They found that the observational features are best explained by a jet from the companion, likely powered by accretion, interacting with the wind of the primary. Their best-fit model
involves MS secondary of  $0.94M_\odot $ accreting at $ {\dot M}_{\textrm{RR}}\sim 2-5 \times 10^{-5}M_\odot$/yr.  The authors appeal  to  Roche lobe overflow, as the eccentric orbit of the secondary  transits through this radius. And because their  accretion rate comes from  inferred disk luminosity and spectra rather than from  jet kinematics, their accretion rate is understandably larger  than 
our minimum   as we shall see.  

\section{Accretion rate constraints}
\label{sec:constraints}

\subsection{Selected Objects}
\label{sec:objs}

Table 1
  shows the 18 objects  from \citet{ppne} 
 and 1 object from \cite{sahai08}  which have fast bipolar outflows and for  which we could 
extract the  jet momentun $p_j=M_{\textrm{j,ob}}v_{\textrm{j,ob}}$ and $t_{\textrm{acc}}$  for use in Eq. (\ref{0.7}). 
\begin{table}
\begin{minipage}{0.5\textwidth}
\caption{List of all the objects we used, their scalar momenta ($p_{\textrm{jet}}$) and acceleration  time scales ($t_{\textrm{age}}$) used.  The first object is from Sahai et al. (2008) , and the rest are from Bujarrabal et al. (2001).}
\begin{tabular}{@{}l c c c}
\hline
Object  & $p_{\textrm{j}}\mathrm{\;[g\:cm/s]}$ & $t_{\textrm{acc}} {\rm or}\  t_{\textrm{ppn}}\mathrm{\;[yr]}$ \\
\hline
IRAS 05506+2414* &  $8.6\times10^{38}$ &$185$\\
IRAS 04296+3429* &  $3.3\times10^{37}$ &$370$\\
CRL 618*  & $8.4\times10^{38}$&$110$ \\
Frosty Leo  & $9.0\times 10^{39}$ &$500$\\
IRAS 17436+5003* & $6.1\times 10^{38}$&$2400$ \\
AFGL 2343*  & $2.8\times 10^{40}$&$1100$ \\
IRC+10420*  & $1.5\times 10^{40}$&$900$ \\
IRAS 19500-1709* & $5.3\times 10^{37}$&$120$ \\
CRL 2688*  & $9.6\times 10^{38}$&$120$ \\
NGC 7027*  & $3.7\times 10^{38}$&$200$ \\
M 2-56 &  $1.3\times 10^{39}$ &$300$\\
Red Rectangle*  & $1.5\times 10^{35}$&$120$ \\
OH 231.8+4.2  & $3.9\times 10^{39}$&$160$ \\
Roberts 22*  & $2.2\times 10^{38}$ &$440$\\
HD 101584  & $1.5\times 10^{39}$&$30$ \\
He 2-113*  & $4.1\times 10^{37}$&$140$ \\
CPD-568032* & $6.1\times 10^{38}$&$140$ \\
M 1-92 & $3.0\times 10^{39}$&$100$ \\
IRAS 21282+5050*  & $5.8\times 10^{38}$ &$1400$\\
\hline
\end{tabular}

\medskip
*: Used $t_{\textrm{ppn}}$ as an upper limit for  $t_{\textrm{acc}}$.

\end{minipage}
\label{tab:objs}
\end{table}

For  $t_{\textrm{acc}}$ in Eq. (\ref{0.7}), we   want the jet acceleration time scale or  the time scale 
that most of the observed jet mass is ejected via the  accretion engine. This  can be at most,  the inferred dynamical age $t_{\textrm{ppn}}$ of the PPN.
For many of the  objects, an acceleration time scale distinct from $t_{\textrm{ppn}}$ is unavailable 
so we use $t_{\textrm{ppn}}$ as an upper limit.
 Since  $t_{\textrm{acc}}\le t_{\textrm{ppn}}$,  use of $t_{\textrm{ppn}}$ in Eq. (\ref{0.7})  would imply a lower limit for the momentum and thus a lower limit on the  minimum required accretion rate.
The objects marked with a "*"  in Table 1
  are those for which  $t_{\textrm{ppn}}$ was used.



\subsection{Graphical representation}
\label{sec:accrate}

\begin{figure*}
\begin{minipage}{0.55\textwidth}
\centering
\includegraphics[width=\textwidth]{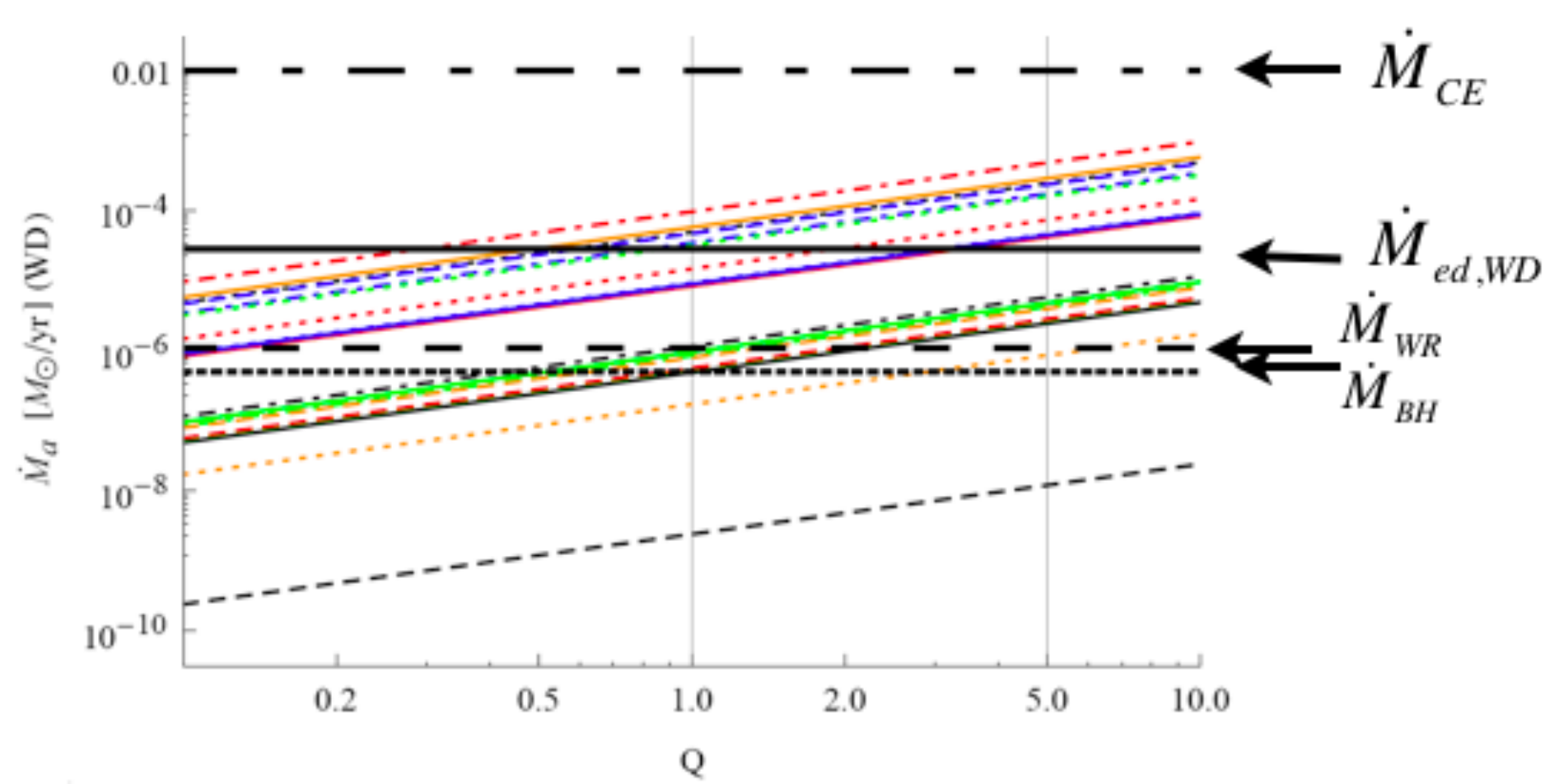}
\end{minipage}
\begin{minipage}{0.55\textwidth}
\centering
\includegraphics[width=\textwidth]{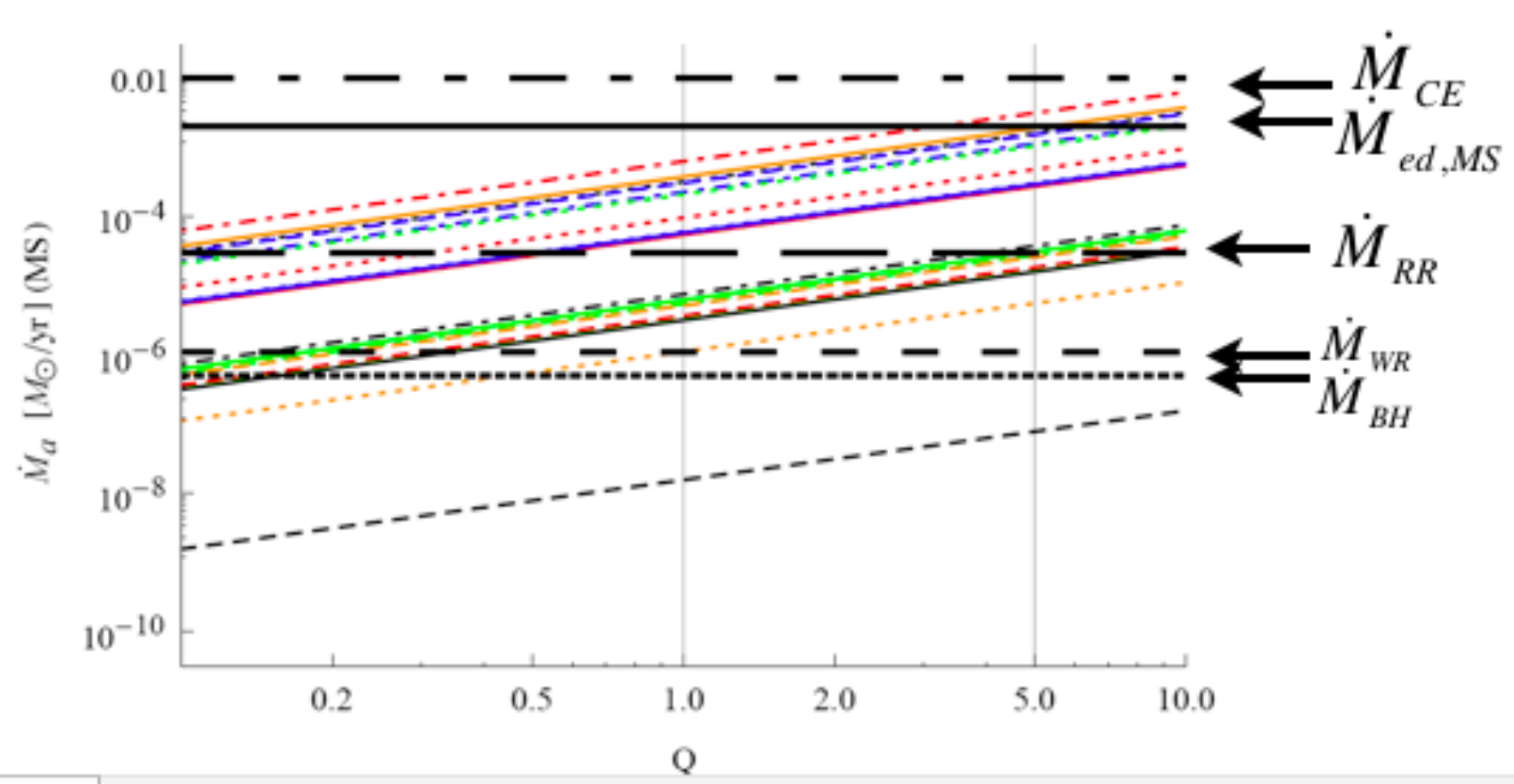}
\end{minipage}
\begin{minipage}{0.35\textwidth}
\centering
\includegraphics[width=\textwidth]{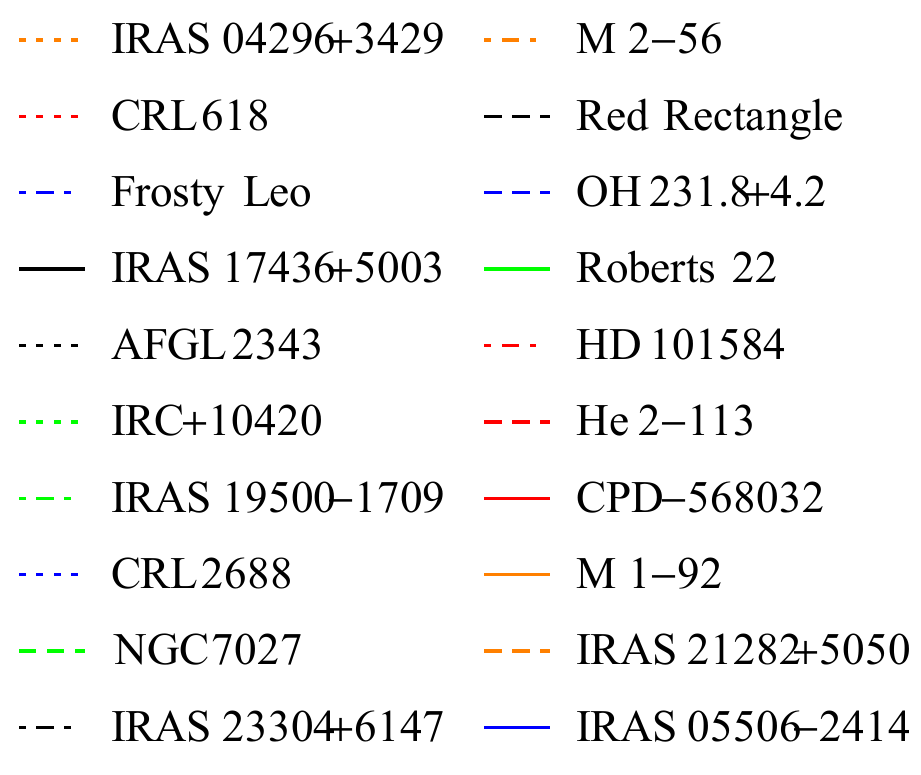}
\end{minipage}
\caption{Mass accretion rate onto the secondary versus $Q$ from Eq. (\ref{0.7}) for the objects of Table 1 for (a)  a $0.6M_\odot$  WD accretor, and (b)  for a  $1M_\odot$ MS accretor. The vertical gridlines show a fiducial range of $Q$ which bounds standard theoretical jet models. The  horizontal lines are indicated by the specific accretion rates discussed in Sec. 4.2. 
Specifically,  $M_{CE}$ refers to common envelope accretion from \citet{rt} ; ${\dot M}_{ed,WD}$ and ${\dot M}_{ed,MS}$ are the Eddington accretion rates for WD and MS stars respectively;  ${\dot M}_{WR}$  is wind Roche lobe overflow for Mira parameters from \citet{mohamed};
${\dot M}_{RR}$ is the Red Rectangle  inferred Roche lobe overflow rate from \citet{wittrr};  ${\dot M}_{BH}$ is the Bondi-Hoyle-Lyttleton rate for specific parameters given below Eq. (\ref{BH}).}

\label{fig:mdotaccmomentum}
\end{figure*}

Using Eq. \ref{0.7} and the  data of Table 1,
 we can  plot  the  minimum required accretion rates for each object as a function of $Q$.  The results are shown as diagonal lines in Fig. \ref{fig:mdotaccmomentum}a for a MS accretor  and in 
 Fig. \ref{fig:mdotaccmomentum}b for a WD accretor. The key difference being that for a WD of mass $0.6M_\odot$ and radius $10^9$cm, 
 the ratio of $(R_a/M_a)^{1/2}$ in Eq. (\ref{0.7}) is  6.5 times smaller than for a MS 1$M_\odot$ star of radius $R_\odot$.


 The horizontal  lines on each plot represent  specific  accretion rate values obtained from the  models of Sec. 3.
In Fig. \ref{fig:mdotaccmomentum}a from top to bottom these lines are:
 ${\dot M}_{CE}
   = 10^{-2}M_\odot$/yr from section 3.3; 
${\dot M}_{\textrm{ED}}=2.9\times 10^{-5}M_\odot$/yr  for a  WD from Eq. (\ref{eq:eddington});   $\dot M_{\textrm{BH}}=1.1 \times 10^{-6}M_\odot$/yr for from Sec 3.1; and  ${\dot M}_{\textrm{WR}}=5\times 10^{-7}M_\odot$/yr from Sec 3.2.
In Fig. \ref{fig:mdotaccmomentum}b from top to bottom these lines are:  ${\dot M}_{CE,S}
   \sim 10^{-2}M_\odot$/yr from section 3.3;  ${\dot M}_{\textrm{ED}}=2\times 10^{-3}M_\odot$/yr  for a $1M_\odot$ MS star from Eq. (\ref{eq:eddington});  
 ${\dot M}_{\textrm{RR}}=(3 \times 10^{-5}M_\odot$/yr) from Sec. 3.4; $\dot M_{\textrm{BH}}=1.1 \times 10^{-6}M_\odot$/yr for from Sec 3.;  ${\dot M}_{\textrm{WR}}=5\times 10^{-7}M_\odot$/yr from Sec 3.2.


Other than the line for  ${\dot M}_{\textrm{RR}}$ from \cite{wittrr}, 
 the   horizontal lines derive from accretion  calculations that depend only on the accretor mass without resolving its radius.
 Since $0.6\le M_\textrm{a}  \le 1.05M_\odot$ in all cases considered, we have used  most of the same horizontal  lines on both plots.



\subsection{Ruling out  modes of accretion }

In Fig. \ref{fig:mdotaccmomentum}a \& b, each diagonal line is a specific object.  For a range of $Q$, all points on  a given diagonal line  above a given horizontal line correspond to the range of accretion rates which cannot be explained by  the model associated with the   horizontal line.


The RR is  the bottom diagonal line in both Fig. 1a \&b but,  as discussed in Sec. 3.4, it is  best modeled by the much  larger Roche overflow onto a MS companion \citep{wittrr}, shown as the ${\dot M}_{\textrm{RR}}$ line in  Fig. (\ref{fig:mdotaccmomentum})b.
This highlights that the actual needed accretion rates can be much higher than the  minima derived from outflow momenta and that other tighter constraints can obviate the need for a lower limit for a given object. We thus focus on non-RR objects.

For  non-RR objects,  all   diagonal lines   lie completely  above the ${\dot M}_{\textrm{BH}}$  in the fiducial range  of $1\le Q\le 5$ for the MS  accretor case (Fig. 1b), 
 ruling out BHL for this $Q$ range. For WD accretors,  BHL is  similarly ruled out for all non-RR objects with $1\le Q\le 5$  except for IRAS 04296+3429 which is ruled out for $Q> 2$.
For a MS star,   Fig. 1b shows that ${\dot M}_{\textrm{WR}}$ is also ruled out for all non-RR objects
 For the WD case of Fig. 1a, ${\dot M}_{\textrm{WR}}$  is acceptable for  IRAS 04296+3429
for $1\le Q\le 5$ and  three other objects for $1\le Q < 2$ but  ruled out for all others.
Fig. 1b also shows that  the RR-RLOF  value ${\dot M}_{RR}$  can  accommodate 7/19 objects.  
A mode of RLOF accretion with accretion rate  significantly larger than that of RR-RLOF  is not ruled out and could accommodate more objects.

Only  the ${\dot M}_{\textrm{CE}}$  line  lies above the diagonal
curves for all 19 objects in both plots.  Sub-Eddington rates for a MS accretor (Fig. 1b) would be  acceptable for all objects in most of the range of  $1\le Q < 5$,  but 8/19 would  require super-Eddington rates for a WD (Fig. 1a) over most of this range. 
\citet{soker04} points out that  for CE accretion modes that involve a shredded companion accreting onto a primary core, a collimated jet may more easily bore through the envelope than for the case in which the an orbiting accretor within the CE is accreting from the envelope  
(as in \cite{rt}).  In this respect,  a higher-than  RR-RLOF mode may be more desirable, in order to reduce
envelope interference with jet collimation. More work is needed.

The extent to which the sources  of Table 1
 are kinematically typical awaits further comprehensive surveys. We do not yet  have an observed distribution function of  PPN fast outflows as a function of  outflow momenta.  In addition,  simulations of the accretion modes discussed in Sec. 3  from which we have derived theoretical constraints are few, and presently for only a limited (albeit reasonable) parameter space of masses and binary radii.  CE simulations  \citep{rt,passy12} are at a particularly nascent state. More theoretical and computational work focused on constraining accretion rates  and developing scaling relations  for each mechanism as a function of mass and binary radius would be
 highly desirable.

\section{Conclusions}
We have shown how  observed  outflow momenta can be used to constrain the viable  accretion engines for bipolar outflows of  PPN. This was accomplished by determining  the minimum needed accretion rates for specific  objects and then  comparing these rates with those derived or computed from specific theoretical engine models.  The main purpose was to demonstrate the method, anticipating that inclusion of more data and more theoretical model accretion rate calculations will  further constrain  allowed paradigms. 

For the PPN sample of 19 objects,  the present constraints (combined with independent constraints for the RR \citet{wittrr})  rule out   BHL  accretion onto a MS or WD star for all but maybe 1 object, and only if the accretor is a WD. For a MS accretor, we can also  rule out M-WRLOF  for all 19 objects when the  accretor is a MS star,  and for 15 objects when the accretor is a WD.
For an MS accretor,  the Red Rectangle Roche lobe overflow  rate can  accommodate 7/19 objects.
Accretion  from within CE evolution can accommodate all 19 objects.  Sub-eddington rates for a MS accretor are acceptable, but 8/19 would require super-Eddington rates for a WD accretor. More work on the latter mode of accretion is desirable, both for accretion onto
the primary core, or onto a CE companion.  For the subset of modes of CE that involve accretion onto an orbiting companion,
the jet may have a tougher time remaining collimated \citep{soker04}  compared to a higher-than Red Rectangle RLOF  rate, and the latter is not  ruled out.

Presently,  jets in only a few percent of known PPN have been studied
and the overall fraction of PPN with jets, let alone the distribution of momenta among them is not well constrained.  The potentially provocative utility of the present approach  helps to motivate more PPN kinematic survey data, more binary searches,   and more theoretical/numerical simulation constraints on accretion rates.  Scaling relations that would  allow the addition of  binary masses and radii as additional constraint dimensions on  plots such as those of Fig. 1 will be particularly useful.


\section*{Acknowledgments} 
  We acknowledge  support from   NSF Grant AST-1109285. 
  We thank N. Soker for  constructive useful comments,  and acknowledge  discussions with
  O. de Marco, A. Frank,  M. Huart\'e-Espinosa, J, Nordhaus, P. Huggins, and R. Sahai.

\bsp

\label{lastpage}

\end{document}